# A Pathway to High $T_c$ and $J_c$ With Hetero-Nano Interface of Tl-2223/Al Heterostructure

**S. H. Pawar*, P.M. Shirage and D. D. Shivagan**

**School of Energy Studies,**
**Department of Physics,**
**Shivaji University, Kolhapur-416 004. INDIA**

## ABSTRACT:

A pathway to high *Tc* and *Jc* with superconducting hetero-nano structures of Tl-2223/Al are reported first time. Two dimensional nano-stripe structure at the interface of Tl-2223/Al was fabricated with room temperature electrochemical complexing technique. The confinement of the charge transport along the hetero-nano stripes at the interface was studied at different temperatures ranging from 300 K to 77 K. The high *Jc* along the interface at different temperatures is explained on the basis of Andreev reflections from the interface followed by confinement of charge transport along *ab*-plane (superconducting path) of Tl-2223 unit cell. The model has been proposed to explain the mechanism of high *Tc* and *Jc* observed at hetero-nano interface. The phenomenon of a pathway to high *Tc* and *Jc* with Tl-2223/Al hetero-nano interface was ascertained by destroying the interface by irradiating the heterostructure.

* Author for correspondence

**shpawar_phy@unishivaji.ac.in**

**pawar_s_h@yahoo.com**

**INTRODUCTION:**

The low dimensional structures (LDS) confined to atomic level has formed a major new branch of physics research giving rise to exciting new physical, optical and electronic properties (*1*). In these LDS, the movement of charge carriers is constrained by potential barriers. The system becomes two, one or zero-dimensional depending on whether the potential barrier confines the carriers in one, two or three dimensions called quantum layers, wires and dots respectively. In the present investigation we have developed first time the quantum layers at superconductor- metal Tl-2223/Ag interface called hetero-nano interface (HNI).

The fabrication of HNI structure with HTc superconductors has many challenges to overcome. At the first instance, the HTc superconductors have anisotropic properties due to their orthorhombic/tetragonal structure having the particular sequential arrangements of different atomic planes in a unit cell. The unit cells of HTc superconductors have to be aligned in thin films with their *c*-axis perpendicular to the plane of the substrate. It needs the perfect *in-situ* monitoring while depositing the films. Secondly, the HTc superconductors are very sensitive to their oxygen stoichiometry. The films need to be treated at high temperatures for their oxidation. The high temperature treatment restricts the formation of abrupt junction between the metal and superconductors as the metal atoms diffuse at the interface forming the mixed nasty phases rather than a junction.

In the present investigation, the authors have developed electrochemical complexing technique to deposit alloyed films of Tl-Ba-Ca-Cu followed by room temperature electrochemical oxidation and obtained single phase $Tl_2Ba_2Ca_2Cu_3O_{10}$ system on Al film with hetero-nano interface. High *Tc* and *Jc* are observed at elevated temperatures and explained on the basis of theoretical model consisting of hetero-nano interface at superconductor and metal junction.

**EXPERIMENTAL:**

Two dimensional nano-stripes structure at the interface between superconductor-metal hetero-nano structure has been formed with room temperature synthesis of HTc superconductor by electrochemical complexing technique developed by Pawar *et. al.* (*2-6*). Aluminum thin films were deposited on amorphous non-conducting glass by vacuum evaporation technique. These films were used as the cathodic substrates in electrochemical copmlexing bath to deposit Tl-Ba-Ca-Cu alloyed films which were then subsequently oxidized at room temperature by electrolytic oxidation technique to yield single phase Tl-2223 superconducting films. The hetero-nano interface has been formed at the junction between Al and Tl-2223 deposits. The confinement of charges across the hetero-nano stripes were studied using four probe geometry, at six different temperatures ranging between 300 K to 77 K. The charge transport across the hetero-nano stripes was studied by using two probe contacts. The hetero-nano stripes were destroyed by irradiating the interface with red He-Ne laser ( $\lambda$ = 632.8 nm, power = 2 mW). This experiment was performed to compare the charge transport with and without hetero-nano interface.

## RESULTS AND DISCUSSION:

Tl-2223/Al heterostructure with hetero-nano interface has been fabricated by room temperature electrochemical complexing technique. The single high *Tc* phase of Tl-2223 is confirmed by X-ray diffraction as shown in fig.1.

The diffraction peaks were indexed with the tetragonal indices. The presence of (117), (205), (118), $(10\underline{15})$, (200), $(10\underline{17})$, $(00\underline{20})$, $(20\underline{10})$, (219), $(11\underline{20})$, $(10\underline{23})$, $(21\underline{19})$ and (309) peaks in XRD-pattern confirms the formation of thallium 2223 phase samples (*7*). The lattice parameters are calculated and found to be a = 3.857 Å and c = 35.7146 Å. These parameters are the same as reported earlier, for 127 K Tl-2223 superconducting phase. The crystallite size was determined by using Schreer's formula and FWHM and was found to be 35 nm.

The pathway of current through superconductor-metal hetero-nano interface was used to study its electrical transport properties. The four probe contacts were made on the surface of HTc film. The current was passed through outer two contacts and the voltage was measured between inner two contacts. The variation in voltage drop with temperature was measured from the 300 K to 50 K. The variation of resistance with temperature was measured by using standard four probe technique with the passage of constant current 1 mA across outer two contacts and developed voltage across inner two contacts. The variation in resistance with temperature is shown in fig.2 (a). Interestingly, it consists of three distinct regions. Region-I ranging from 300 K to 175 K, is designated as 'bulk resistance' region. This is due to the variation of bulk resistance of $Tl_2Ba_2Ca_2Cu_3O_{10}$ in normal state. The region-II ranging from 175 K to 125 K is designated as interface transport region. In this region there is sudden drop in resistance and current transports in the plane of nano-interface. The region-III below 125 K is due to the variation of resistance of Al film, which is metallic in nature.

The region-II due to HNI is of interest and studied for super pathway of charge transport analogous to *Jc* and *Tc* in superconducting state. The onset of the region-II was found to change with the values of constant current passed through the heterostructure. For 5 mA constant current input it starts at 290 K as shown in fig. 2(b). This reveals that there are different high *Tc* values for super pathway to charge transport in HNI and depends on the magnitude of the input current.

The charge transport in the interface was further studied by measuring the voltage drop, with the increase in input current at various temperatures. The plots recorded at six different temperatures are shown in Fig.3. It is seen that there are two distinct regions called ohmic (at low current) and non-ohmic regions (at high current). For low values of current, the voltage drop increases linearly with increase in current as shown in inset plot of fig.3. The slope of this plot gives the value of the bulk resistance of Tl- 2223 region in heterostructure. The ohms law obeys up to certain values of current and there is sudden increase in V at I = Ic. Beyond this value of current, the voltage drop remains constant for the passage of any large current dis-obeying the ohm's law. This is attributed to the passage of charge carriers through nano-interface giving rise to super high way to current with zero resistance. This results in high value of *Tc*. The plot recorded at 150 K, 100 K and 77 K are overlapped to each other and the values of the V are very small. This may be due to the fact that the bulk region of Tl-2223 in heterostructure goes into superconducting state in this temperature range.

To understand the charge transport across the junction, many workers have studied the interface of superconductor-metal heterostructure both experimentally and theoretically. There are two key processes involved across the junction: Andreev reflection and proximity effect. These phenomena provide interesting transport characteristics. The Andreev reflection affects the charge transport due to tunneling across the junction. Recently, Giubileo *et al.* (*8*) have studied Andreev reflection in $Tl_2Ba_2CaCu_2O_8$/Ag interface measuring tunneling conductance. The tunneling conductance is dependent on the barrier of the interface. The barrier of the interface is highly transparent if the junction is a point contact. There are two phenomena occurring at the interface: reflection and transmission of the charge carriers.

Thus in the point contact junction, there is less reflection and maximum transmission. In the present investigation, we have developed first time the large reflection and less transmission superconductor-metal hetero-nano interface for super pathway to current along the barrier of the interface rather than across the interface. The reflectance of the interface depends on area of the junction, barrier potential and energy gap parameter of superconductor. In addition to the reflection of the charge carrier, we have introduced the concept of the transportation in the plane of the interface. The heterostructure of Tl-2223/Al possesses hetero-nano-interface. The height of the potential parameter is estimated by studying the conductance spectra of hetero-nano interface by applying external voltage across the interface at different temperatures. The typical plot of the conductance spectra at 77 K is shown in fig. 4. The values of the energy gap parameter are found to be very large than that of the pure superconductor. It is found to vary with the temperature of hetero-nano-interface.

The theoretical model has been suggested to explain the mechanism of charge transport at Tl-2223/Al hetero-nano interface. This model consists of hetero-nano interface sandwiched between bottom of bulk Tl-2223 and top of Al metal. The charges, e-, entered into hetero-nano interface are reflected back into the interface from highly strained layer between Al surface and hetero-nano interface with high barrier potential ten times larger than superconductor gap. The reflected charge carriers perpendicular to the interface along the c-axis of Tl-2223 face the high resistive path and they are diverted along the interface through the *ab*-plane (superconducting) of Tl-2223 unit cell. This gives rise to a super pathway to charges along the interface and hence high $J_c$

In order to ascertain the role of hetero-nano interface in observing high *Tc* and *Jc* effect, we have destroyed the interface by irradiating the heterostructure by red He-Ne laser radiations. The plot of the variation in resistance with temperature during laser irradiation is shown in fig.5. It is seen that the heterostructure shows the regular pattern of superconducting behavior of $Tl_2Ba_2Ca_2Cu_3O_{10}$. The variation in voltage drop with current is shown in the inset of fig.5, giving rise to *Jc* values of the order of 2.4 x $10^6$ A/cm$^2$

**Conclusions:**

Hetero-nano interface at Tl-2223/Al junction is successfully developed by a novel electrochemical complexing technique. It consists of hetero-nano interface sandwiched between between Tl-2223/ Al which acts as the super-hiway path for the transport of current. The critical temperature of the supercurrent transport across the interface, with the maximum of 290 K, is found to vary with the magnitude of current; suggesting superconducting nature of the hetero-nano interface at this temperature. The high-energy gap of the order of 220 meV measured by the differential contact conductance suggests the presence of the high potential barrier of nano-interface with maximum reflecting type Andreev states. The theoretical model has been proposed to explain the mechanism of High $T_c$ and $J_c$ observed at hetero-nano interface. The zero resistance transition, after destroying the interface by laser irradiations, at 104 K with the $Jc$ values of 2.4 x $10^6$ A/$cm^2$ reveals the formation device quality single phase Tl-2223 films.

9. We wish to thank University Grants Commission, New Delhi for the financial support under superconductivity R & D programme.

## Figure captions:

Fig.1. X-ray diffraction pattern of Tl-2223/Al

Fig.2 The variation of normalized resistance of Tl-2223/Al at two different temperatures: for (a) 1 mA and (b) 5 mA

Fig.3. Variation of potential drop (V) with the passage of current through Tl-2223/Al heterostructure at T = 300, 250, 200, 150, 100 K, and 77 K.

Fig 4. Conductance spectrum measured at 77 K for Tl-2223/Al heterostructure.

Fig.5 The plot of the variation in resistance with temperature after destroying hetero-nano interface: a) during laser irradiation and b) after laser irradiation.

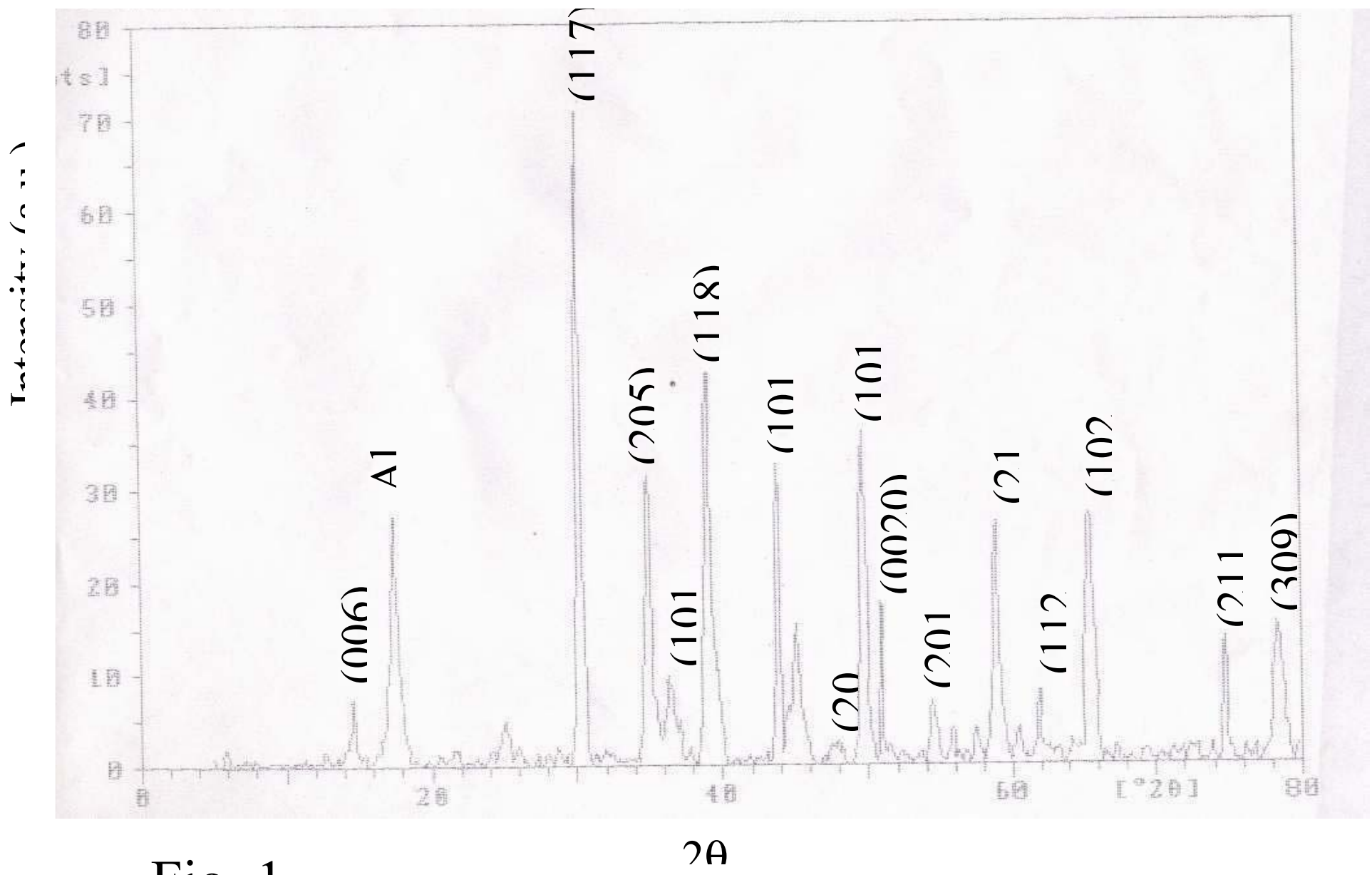

Fig. 1

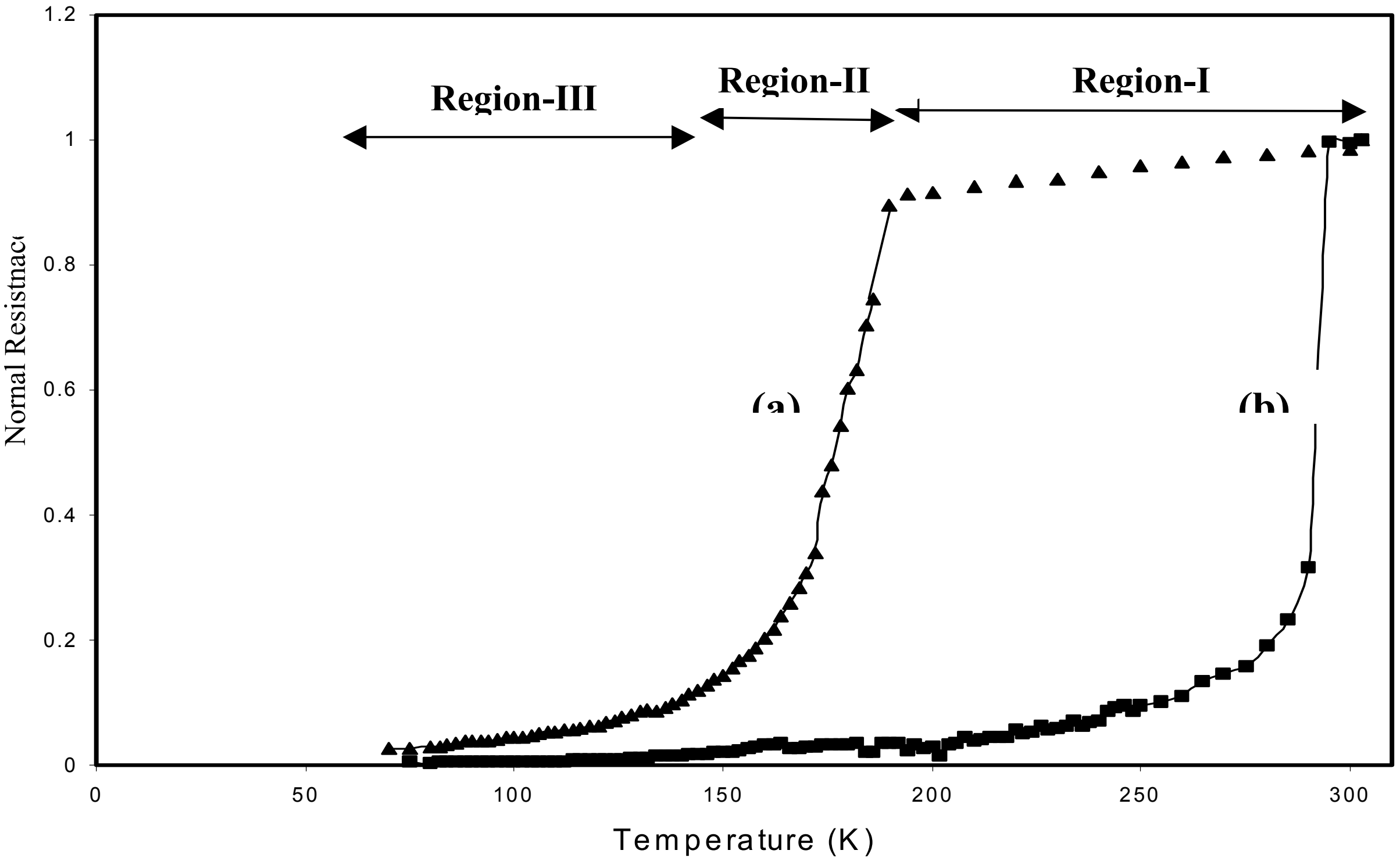

Fig. 2

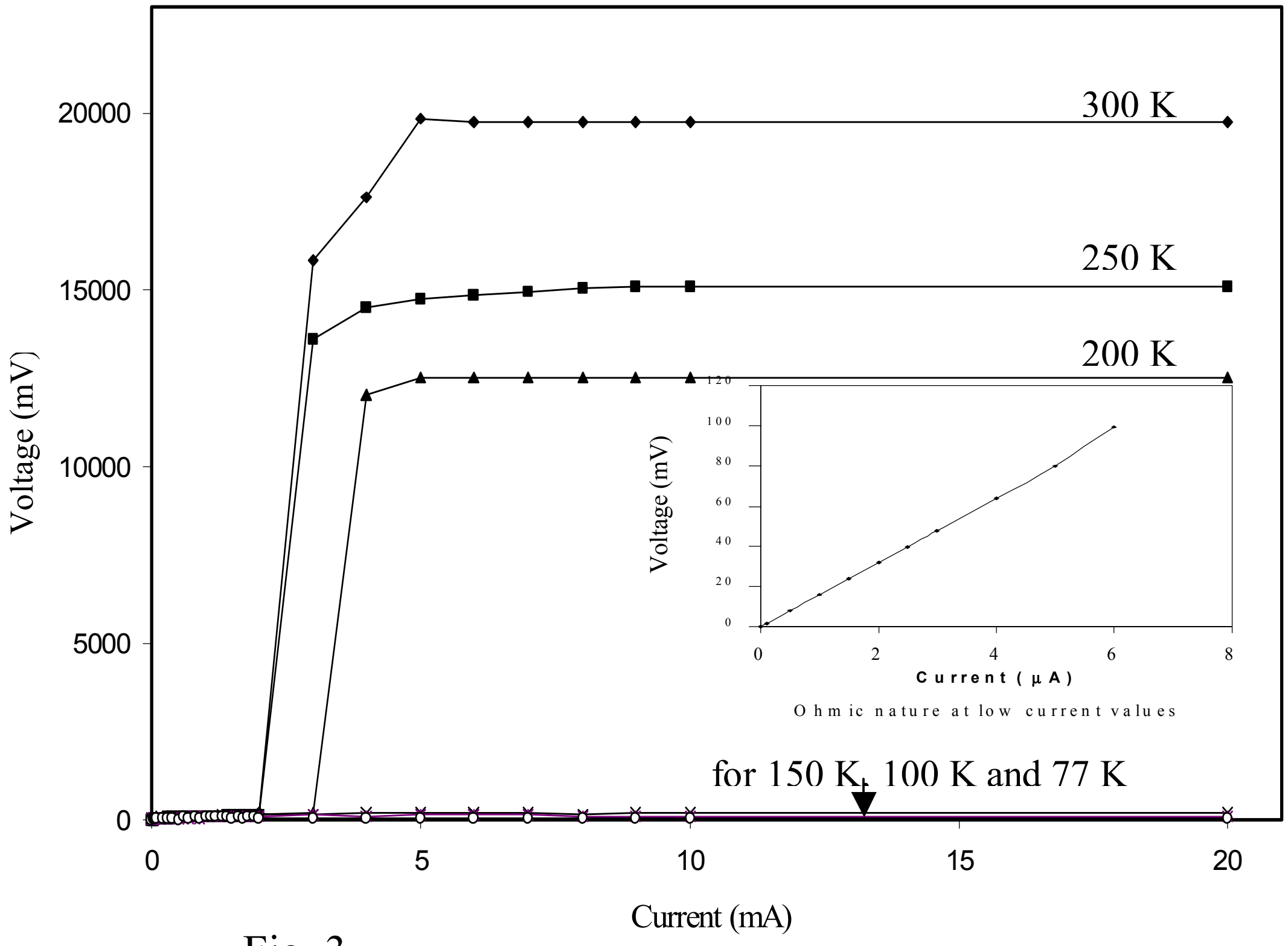

300 K
250 K
200 K
20000
15000
10000
5000
0
Voltage (mV)
0
5
10
15
20
Current (mA)
for 150 K, 100 K and 77 K
120
100
80
60
40
20
0
Voltage (mV)
0
2
4
6
8
Current (μA)
Ohmic nature at low current values

Fig. 3

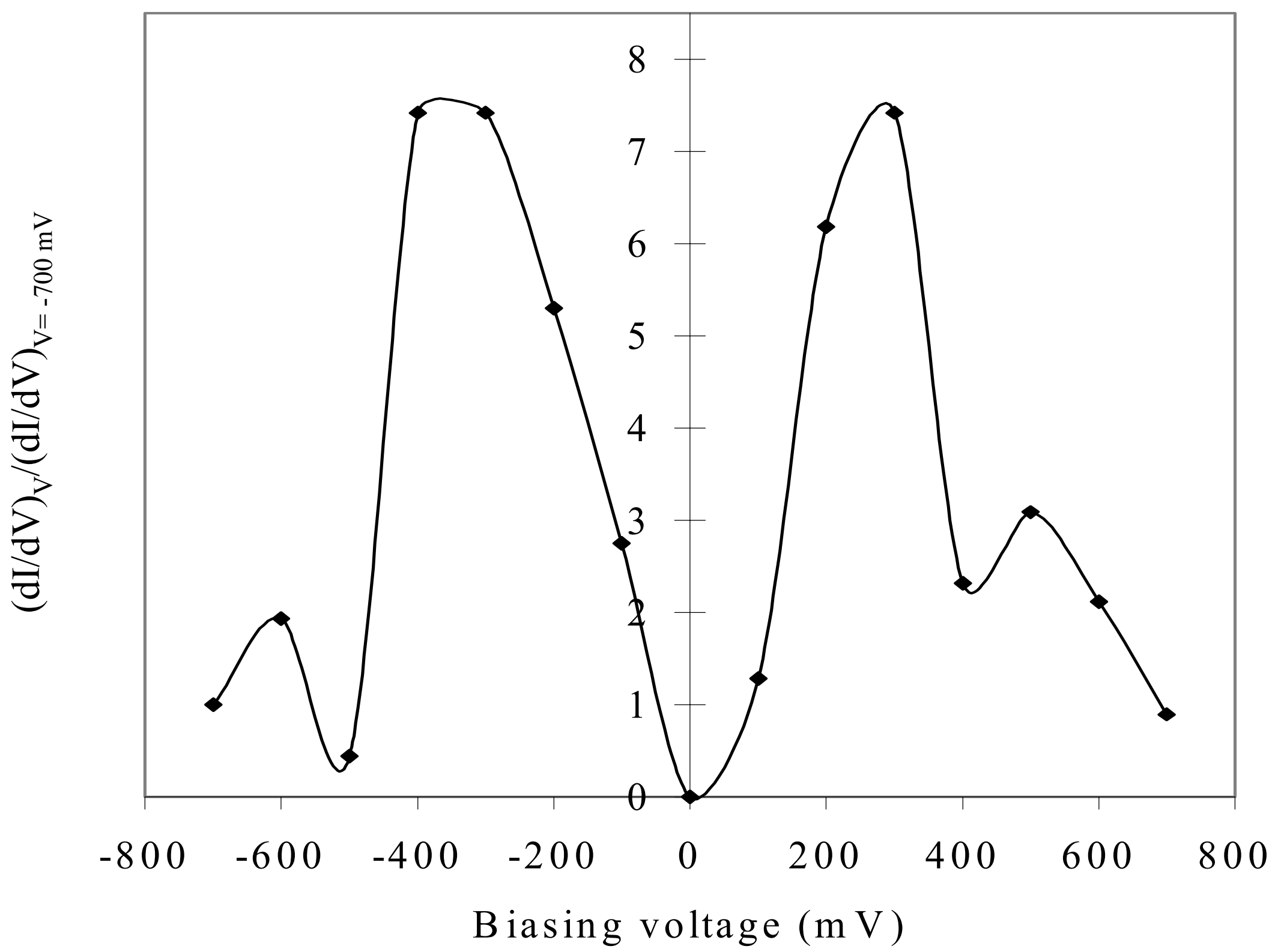

(dI/dV)$_V$/(dI/dV)$_{V= -700\ mV}$
8
7
6
5
4
3
2
1
0
-800 -600 -400 -200 0 200 400 600 800
Biasing voltage (mV)

Fig.4

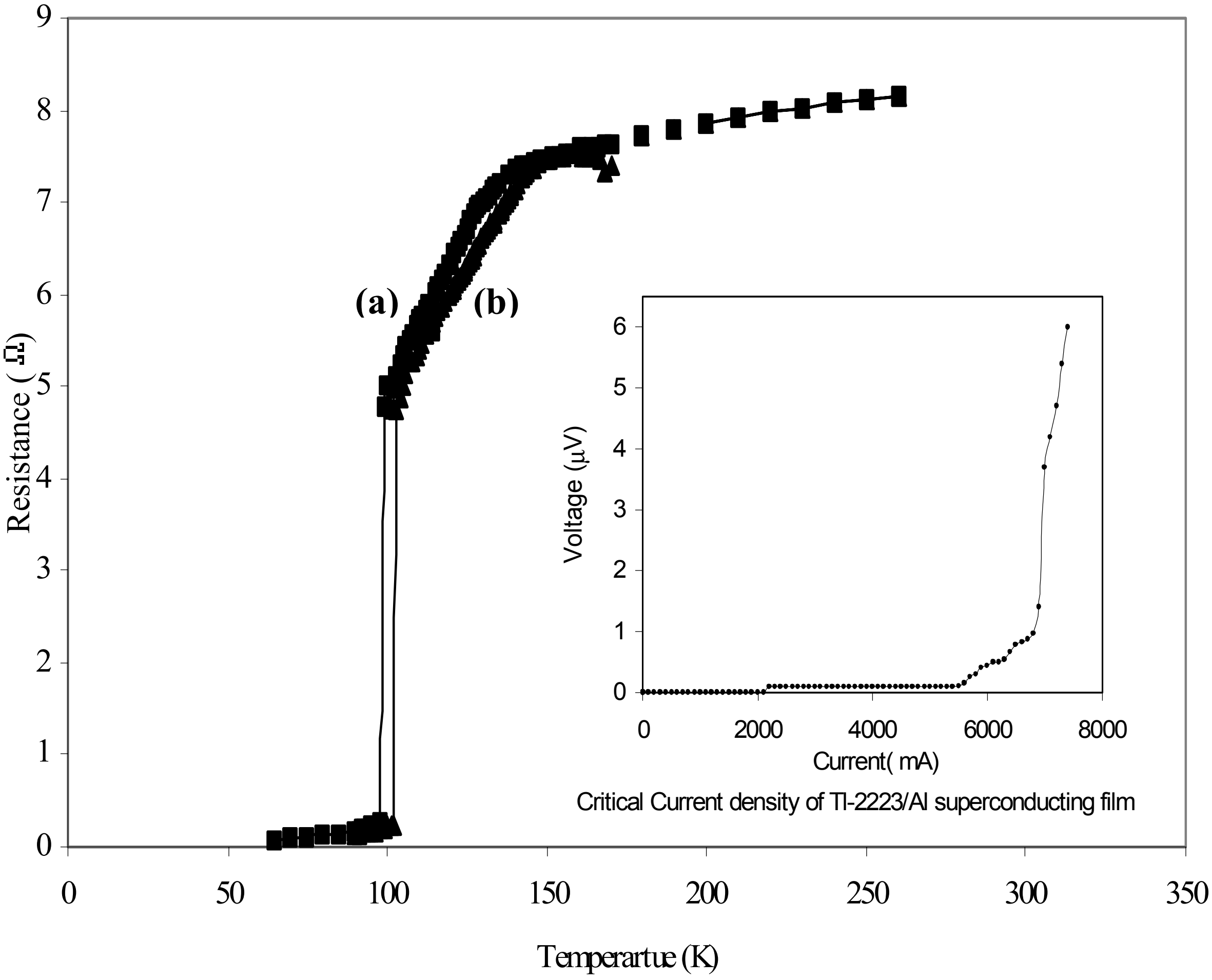

**Fig.5**